%% file: K3-35_Tafoya_etal.tex
\documentclass[nofigure]{pasj00}
\SetRunningHead{Distance to the PN K~3$-$35}{Distance to the PN K~3$-$35}

\usepackage{times}
\usepackage[dvips]{graphicx,color}

\draft

\def\kms{~km~s$^{-1}$}

\def\h2o{H$_{2}$O}

\def\k3{K~3$-$35}
\def\j1925{J192559.6$+$210626}

\begin{document}

\title{Measurement of the Distance and Proper Motions of the H$_{2}$O Masers in the Young Planetary Nebula \k3}
\author{Daniel TAFOYA$^{1}$, Hiroshi IMAI$^{1}$, Yolanda G\'OMEZ$^{2}$, Jos\'e M. TORRELLES$^{3}$, 
Nimesh A. PATEL$^{4}$,  Guillem ANGLADA$^{5}$, Luis F. MIRANDA$^{6,7}$, Mareki HONMA$^{8}$, Tomoya HIROTA$^{8}$, 
Takeshi MIYAJI$^{8}$ 
	}
\affil{$^{1}$Department of Physics and Astronomy, Graduate School of Science and Engineering,\\
	 Kagoshima University, 1-21-35 Korimoto, Kagoshima 890-0065, Japan\\
   	$^{2}$Centro de Radioastronom\'\i a y Astrof\'\i sica, Universidad Nacional Aut\'onoma de M\'exico,\\ 
	58089 Morelia, Michoac\'an, M\'exico\\ 
	$^{3}$Instituto de Ciencias del Espacio (CSIC)-UB/IEEC, Facultat de F\'{\i}sica, Universitat de Barcelona,\\ 
	 Mart\'{\i} i Franqu\`{e}s 1, E-08028 Barcelona, Spain\\
	$^{4}$Harvard-Smithsonian Center for Astrophysics, 60 Garden Street, Cambridge, MA 02138, USA\\
	$^{5}$Instituto de Astrof\'{\i}sica de Andaluc\'{\i}a, CSIC, Apartado Correos 3004, E-18080 Granada, Spain.\\
	$^{6}$Instituto de Astrof\'{\i}sica de Andaluc\'{\i}a - CSIC, C/ Glorieta de la Astronom\'{\i}a s/n, E-18008 Granada, Spain\\
	$^{7}$Departamento de F\'{\i}sica Aplicada, Facultade de Ciencias, Universidade de Vigo\\
	 E-36310 Vigo, Spain (present address)\\
	$^{8}$Mizusawa VLBI Observatory, National Astronomical Observatory of Japan
	}

\email{dtafoya@milkiway.kagoshima-u.ac.jp}

\maketitle

\begin{abstract}
In this paper we present the results of very long baseline interferometry (VLBI) observations carried 
out with the VLBI Exploration of Radio Astrometry (VERA) array and the Very Long Baseline Array (VLBA) 
toward H$_{2}$O masers in a young planetary nebula \k3. From the VERA observations we measured the 
annual parallax and proper motion of a bright water maser spot in \k3. The resulting 
distance is $D=$~3.9$^{+0.7}_{-0.5}$~kpc. This is the first time that the parallax of a 
planetary nebula is obtained by observations of its maser emission. On the other hand, the proper motion of 
\k3 as a whole was estimated to be $\mu_{\alpha}=~-$3.34$\pm$0.10~mas~yr$^{-1}$, $\mu_{\delta}=~-$5.93$\pm$0.07~mas~yr$^{-1}$. 
From these results we determined the position and velocity of \k3 in Galactic cylindrical coordinates: 
($R,\theta,z$)~$=$~(7.11$^{+0.08}_{-0.06}$ kpc, 27$\pm$5$^{\circ}$, 140$^{+25}_{-18}$ pc) and 
($V_{R}, V_{\theta}, V_{z}$)~$=$~(33$\pm$16, 233$\pm$11, 11$\pm$2) \kms, respectively. Additionally, 
from our VLBA observations we measured the relative proper motions among the water maser spots located in 
the central region of the nebula, which have been proposed to be tracing a toroidal structure. The distribution 
and relative proper motions of the masers, compared with previous reported observed epochs, suggest that such 
structure could be totally destroyed within a few years, due to the action of high velocity winds and the expansion 
of the ionization front in the nebula. 

\end{abstract}

\section{Introduction}

Determining the distance to astronomical sources represents one of the most challenging tasks in modern 
astronomy. The measurements become more difficult in the cases where there is no standard candle or ruler 
that establishes a direct relationship between an observable parameter and the distance. This is the 
situation of planetary nebulae (PNe) for which our knowledge on the distance still remains poor. The 
distance is a crucial parameter for studying the physical conditions and evolution of PNe. Furthermore, 
it also provides information on the location and velocity of these objects within the Galaxy, which 
indicate the stellar population to which PNe belong and their kinematical history in the Galaxy 
(see e.g. Imai et al. 2007). However, until now, the distances to less than 50 PNe have been estimated 
individually with a reasonable accuracy (e.g. Guzm\'an et al. 2009 and references therein). While most 
of those measurements were performed through indirect methods that are based on assumptions that cannot 
hold true in all cases, only about one third of them have been obtained by a direct method: trigonometric 
parallax of the central star. This method demands very precise astrometry (of the order of milliarcsecs) 
of a source at several epochs, resulting in a very arduous task. Fortunately, in recent years the techniques 
of VLBI in radio astronomy have been improved in such a way that, by carrying out careful astrometric 
observations, we can measure the trigonometric parallax, and subsequently the distance to the cosmic objects 
with great accuracy (e.g. Imai et al. 2007; Loinard et al. 2007, Nakagawa et al. 2008; Reid et al. 2009; 
Moellenbrock et al. 2009, Sato et al. 2010). An important constraint on these observations is that the 
sources must exhibit emission with high surface brightness. This requirement is met by the maser emission. 
Therefore, we can accurately determine the distance to PNe that present this type of emission by observing 
the masers in their envelopes. Moreover, these observations can be used as a powerful tool to study the 
kinematics of the circumstellar gas in water maser emitting PNe at very small scale.

\k3 is the first PN for which the association with water maser emission was discovered (Miranda et al. 
2001). Since then, surveys of water masers toward very young PNe have been carried out, aiming to find more 
sources of the same type. As a result, two additional PNe have been found to be harboring water molecules: 
IRAS 17347-3139 (de~Gregorio-Monsalvo et al. 2004) and IRAS 18061-2505 (G\'omez et al. 2008). These three 
sources are characterized by showing bipolar lobes and a narrow and obscured equatorial band, as seen in 
the optical and infrared images (Miranda et al. 1998; Sahai et al. 2007). Thus, it has been suggested that 
a bipolar wind and a high density equatorial torus could be present in these PNe (G\'omez et al. 2008; 
Tafoya et al. 2009), similarly to those observed in their progenitors, the pre-planetary nebulae. The 
presence of dust and high density gas could explain why the water maser molecules remain in the nebula even 
when the central star is already emitting a large amount of ionizing photons (Tafoya et al. 2007).  

Miranda et al. (2001) observed the water masers in \k3 with the Very Large Array (VLA) and found that they 
are located in the central region of the nebula as well as at the tips of the two point-symmetric radio 
lobes of the nebula. The masers at the tips of the radio lobes resemble those of the so-called {\it water 
fountain nebulae} (Imai et al. 2002; Claussen et al. 2009), and they seem to be associated with a bipolar, 
precessing wind. On the other hand, the masers in the central region appear distributed along the equatorial, 
dark band, suggesting that they are associated with an equatorial torus. This idea has been considered by 
Uscanga et al. (2008), who modeled the central masers of \k3 as arising from a rotating-expanding ring-like 
structure. However, detailed measurements of the internal motion of these masers are necessary to better 
understand the kinematics of the different structures in this nebula.

In order to determine with higher accuracy the distance to the young planetary nebula \k3, and to measure 
the proper and internal motions of the water masers in the nebula, we carried out multi epoch VLBI 
observations. For this purpose, we have used the VLBI Exploration of Radio Astrometry (VERA) Array of the 
National Astronomical Observatory of Japan (Kobayashi et al. 2003) and the Very Long Baseline Array 
(VLBA) of the NRAO\footnote[9]{The National Radio Astronomy 
Observatory is a facility of the National Science Foundation operated under cooperative agreement by 
Associated Universities, Inc.}. The details of the observations are described in \S 2. The annual parallax 
and proper motion measurement of \k3, as well as the internal motions of the masers are presented in \S 3. 
In \S 4 we discuss the results and their implications on the study of \k3. 

\section{Observations and data reduction}

\subsection{VERA Observations}

The observations of the \k3 \h2o masers at $\sim$22 GHz with VERA were carried out during 16 epochs from 2007 
October to 2009 August. Table 1 gives a summary of these observations. At each epoch, the total 
observation time was about 8 hours, including the scans on \k3 and the fringe phase and position reference 
source, J192559.6$+$210626 (hereafter J1925+2106), which is one of the International Celestial Reference 
Frame (ICRF) sources. Since these two sources are spatially separated by only 0\arcdeg.57 it was possible to 
use the VERA's dual-beam system. 
The telescope observed \k3 and J1925+2106 for 8 minutes every 20 minutes; in between, the target maser source 
IRAS~19312+1950 and its correspondent reference source were also observed. The astrometric results for that 
source are presented by Imai et al. (2010). 
The signals were digitized in four quantization levels, and then divided into 16 base-band channels (BBCs) in a 
digital filter unit, each of which had a bandwidth of 16 MHz. One of the BBCs was assigned to the frequency of 
the \h2o maser emission in \k3 while the other 15 BBCs were assigned to the continuum emission from J1925+2106.

The data correlation was made with the Mitaka FX correlator. The accumulation period of the correlation was set 
to 1 second. The correlation outputs consisted of 512 and 32 spectral channels for the \h2o maser and reference 
continuum emission, respectively. A velocity spacing of 0.42\kms~ was obtained in each spectral channel for the 
\h2o maser emission. 

The data reduction was mainly made with the Astronomical Imaging Processing System (AIPS) package of the NRAO. 
To achieve a good astrometry, we performed the following procedure. At first, we recalculated the delay-tracking 
solutions for the correlated data using an improved delay-tracking model. Through the whole data analysis, we 
adopted the coordinates of the delay-tracking center: 
$\alpha_{J2000}=$ 19$^{\mbox h}$27$^{\mbox m}$44$^{\mbox s}.$023,  
$\delta_{J2000}=~+~\!\!$21$^{\circ}$30$^{\prime}$03$\arcsec.$44 for \k3 and 
$\alpha_{J2000}=$ 19$^{\mbox h}$25$^{\mbox m}$59$^{\mbox s}.$605352,  
$\delta_{J2000}=~+~\!\!$21$^{\circ}$06$^{\prime}$26$\arcsec$.16198 for the position reference source J1925+2106. 
The delay-tracking solutions included delay contributions from the atmosphere, which were estimated using the 
global positioning system data (Honma et al. 2008). Subsequently, the differences due to instrumental delays between 
the two signal paths in the dual beam system were calibrated using differential delays, which were measured using 
artificial noise signals injected at the same time into the two receivers. Then, 
fringe-fitting and self-calibration were performed on the fringe phase and position reference source data, whose 
solutions were applied to the maser emission data. Only the solutions of the BBC with the same frequency as that 
of the \h2o maser emission were used for the fringe-phase compensation. Finally, the image cubes of the maser source 
were obtained by visibility deconvolution through the CLEAN algorithm. A few CLEAN boxes were specified after trial imaging. 
This is necessary to avoid false identification of maser emission in the phase-referenced {\it dirty} images, which have 
high-level side lobes due to imperfect phase compensation. The typical rms noise in the emission-free spectral channel 
images is given in Table 1. 

\input{table_1.tex}

\subsection{VLBA Observations}

The VLBA observations were carried out at 3 epochs: 2003 September 24, 2003 November 20, and 2003 December 21. 
The observation settings were basically the same for the 3 epochs. The ten antennas of the VLBA were used in 
all the epochs. \k3 was observed during intervals of 40 minutes, alternated with observations of $\sim$10 minutes 
on the calibrators J1925+2106, BL Lac and 3C345. The total on-source observation time was $\sim$6.5 hours. One IF 
with right circular polarization was used for our measurements. The bandwidth was 8 MHz centered at 20\kms~ with 
respect to the local standard of rest (LSR). The data was correlated with the VLBA correlator in Socorro, NM, 
sampling 256 channels, which resulted in a spectral spacing of 31.25 kHz, corresponding to 0.42 km~s$^{-1}$.

The data reduction was performed using the AIPS package. The amplitude calibration was made using a priori knowledge 
of the gains and system temperatures of the system and antennas. The residual delays were obtained using the 
source J1925+2106, and the bandpass calibration was made by using the sources BL Lac and 3C345. After this calibration, 
we performed fringe-fitting for the spectral channel with the strongest emission. In this process, the information of the 
absolute position is lost. Subsequently, we self-calibrated the visibilities of this channel and the solutions were copied to 
the rest of the spectral data. The self-calibrated data was deconvolved through the CLEAN algorithm; the rms noise in a single 
channel of the final data cube was 
typically $\sim$5~mJy~beam$^{-1}$. Subsequently, we proceeded to search for maser emission using the AIPS task SAD 
with the criteria of detecting emission above 10 times the noise level. From the candidate maser emission, we selected 
only those features that appeared in more than three contiguous spectral channels and found their positions by fitting a 
two-dimensional Gaussian model to the brightness distribution. The relative positional accuracy of the detected maser 
spots was typically better than 10~$\mu$as.\
  
\section{Results}

Thanks to the dual-beam system of VERA we were able to determine the positions of the water masers of \k3 
with respect to the ICRF, which is a quasi-inertial reference frame based on the radio position of 212 
extragalactic sources whose positions are known better than 0.5 mas (Ma et al. 1998).  The absolute position 
error of maser spots with respect to the reference source achieved by our VERA observations was 
$\lesssim$0.1~mas. Therefore, they  were suitable to determine the annual parallax and proper motion of \k3. 
On the other hand, our VLBA observations provided 
only the relative positions of the masers with respect to a reference feature. In consequence,  we did not use them 
for the calculation of the annual parallax and proper motion of \k3 as a whole. However, these observations provided higher 
sensitivity, allowing us to detect fainter maser features than those detected with VERA. Thus, the VLBA observations 
were used to quantify the relative internal motions of the masers. Following we describe the results of the observations 
from each array.  

\subsection{Annual parallax and proper motion of K 3-35}

\input{table_2.tex}

\input{table_3.tex}

Table 2 shows the parameters of the water maser spots (i.e. maser emission in each individual channel map) detected 
with VERA during 14 epochs of observation. We found maser emission in the channels corresponding to the range of 
velocities v$_{\rm LSR}\sim$20.88 -- 23.00\kms. The velocity of these masers corresponds to that of those located in the 
central region of \k3. The masers associated to the bipolar lobes, found by Miranda et al. (2001), are likely to have 
disappeared, given that they were not detected neither by de Gregorio-Monsalvo et al. (2004), nor in our observations. 
When the emission from all the epochs is velocity-integrated and mapped, we identify three maser features, which are 
defined as groups of maser spots that for a given observation epoch are spatially separated by $\lesssim$0.5~mas with a
$\Delta$v $\lesssim$1.3~\kms. These maser features correspond to physical maser clumps and they are shown in the 
left panel of Figure~1, labeled as $A$, $F$ and $H$ (also see Table~2). Notice that for the features $A$ and $F$ we have 
used the same labels as for those detected with the VLBA, shown in Figure~2.  The reason of doing this is that we noticed 
that the relative positions, as well as the 
LSR velocities, of two of the maser features detected with VERA are quite similar to those of the features labeled as 
$A$ and $F$ from the VLBA observations. Therefore, it is reasonable to assume that they are the same maser features.
The feature $H$ was detected in two epochs with VERA but not with the VLBA. Within each feature, the relative
positions of the maser spots change slightly ($\sim$50 $\mu$as) 
from one epoch to another. This can be due to a combination of the uncertainty in the determination of the position and the 
turbulent motion of the gas within the masing region. This variation is significant when compared to the parallax 
($\sim$250 $\mu$as, see below). Therefore, in order to determine the annual parallax from the maser emission, it 
is necessary to trace the motion on the sky of a spot that remains relatively stable in position within the masing 
cloud and that can be detected during most of the epochs of observation. It has been found that this requirement is 
generally met by the brightest spots, i.e. their positions are better determined and they are not significantly affected by the 
turbulent motions within the masing region (e.g., Imai 1999). From the three maser features found in our observations 
with VERA, the feature $F$ was detected in most of the epochs. Within this maser feature, there are 
three spots with velocities at v$_{\rm LSR}\sim$22.15\kms, $\sim$22.57\kms and $\sim$22.99\kms; 
only the one at v$_{\rm LSR}\sim$22.57\kms was detected at sufficient number of epochs to be used for the parallax 
and proper motion measurements.

\begin{figure*}[h!]
  \begin{center}
    \includegraphics[angle=-0,scale=0.5]{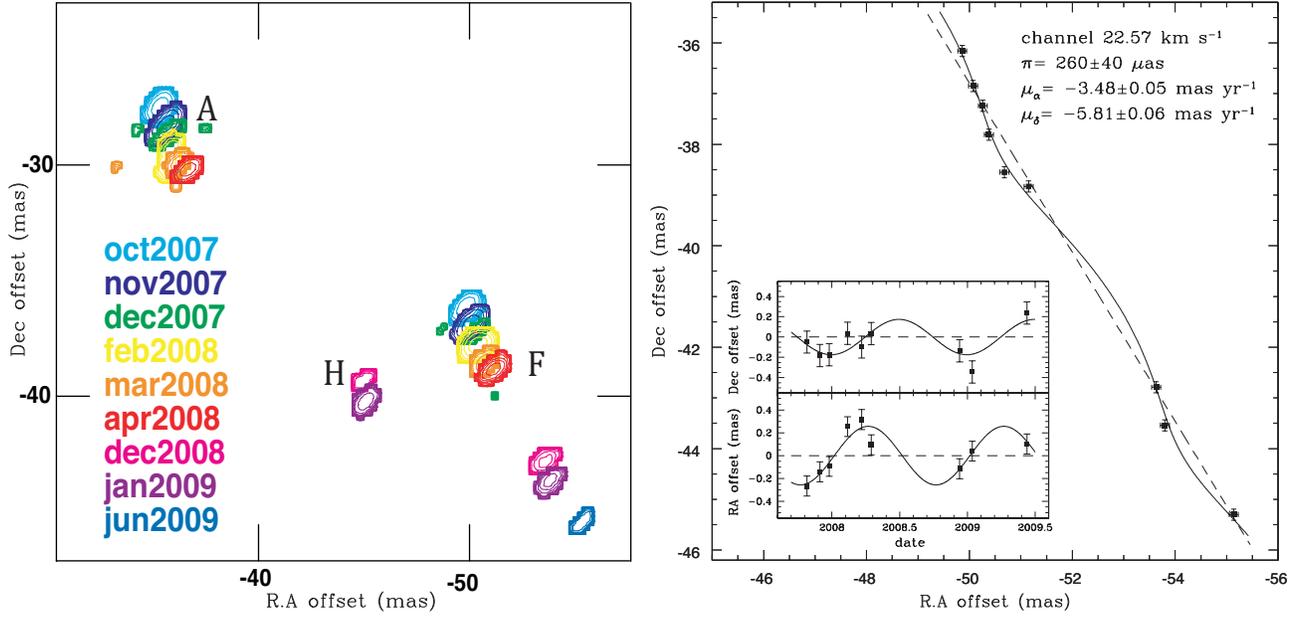}
  \end{center}
  \caption{Left: Maps of the maser emission detected with VERA at the epochs used for the measurement of the annual parallax and 
  proper motion (Table 3). The labels ($A$ and $F$) indicate the same maser features as  those shown in Table 2 and 
  Figure 2.  The first contour indicates the 6-sigma noise level. The following contours show intensity levels at 1\% of peak intensity 
  and they increase by a factor of 1.6. Right: Motion of the water maser spot at v$_{\rm LSR}=$~22.57\kms~(i.e. $m_{\rm 22.57}$) in \k3 and the 
fitted kinematical model consisting of a linear proper motion and a sinusoidal annual parallax  
(dashed line and solid line, respectively). 
The R.A. and Dec. offsets are set with respect to the phase-tracking center of the 22.57\kms\ component.
The inset shows the motion of the spot as a function of time, after removing the linear proper motion 
component. The error bars show the mean standard deviation of the data from the model 
($\sigma_{\rm RA}=$ 0.09 mas, $\sigma_{\rm Dec}=$ 0.11 mas).}
\label{.....}
\end{figure*}

\input{table_4.tex}

The motion on the sky of the maser spot at v$_{\rm LSR}\sim$22.57\kms~($m_{\rm 22.57}$) was modeled as a 
combination of a linear proper motion and the sinusoidal component due to the annual motion of the Earth. Thus, any 
other non-linear proper motion was considered to be negligible. In a first approach, we fitted the data from all the observation 
epochs. We noticed that some data points showed a significant offset from the model, as compared to the rest of the 
observations. We also noticed that the same offsets were found in the data of the source IRAS~19312+1950 (see \S 2.1). 
This source was observed during the same observation runs as \k3 and the results of the observations are presented 
by Imai et al. (2010). Since these two sources and their corresponding calibrators are different and independent from 
each other, we concluded that the observed offsets of some data points from the model must be due to a systematic 
problem during the observations at those epochs, which could not be 
removed during the calibration process. Therefore, we {\it flagged out} the points corresponding 
to the epochs that showed the systematic offset and performed the fitting again. The last 
column of Table~1 indicates whether the data was used for the fitting or not; Table 3 shows 
the parameters of the data points used in the fitting process. 

The position of the peak of the emission of $m_{\rm 22.57}$ at the different epochs, along with 
the fitted model are shown in the right panel of Figure~1. In the inset of this Figure, we also show the motion of 
the maser spot as a function of time after subtracting a derived linear proper motion of 
$\mu_{\alpha}=-$3.48~mas~yr$^{-1}$, $\mu_{\delta}=-$5.81~mas~yr$^{-1}$, which corresponds to the absolute 
proper motion of the feature $F$. The error bars indicate the 
mean standard deviation of the data from the model: ($\sigma_{\rm RA}=$~0.09~mas, $\sigma_{\rm Dec}=$~0.11~mas). 
They were set to be the same for all the data points. The annual parallax was estimated to be 260$\pm$40 $\mu$as, 
which corresponds to a distance to the source of 3.9$^{+0.{\bf 7}}_{-0.{\bf 5}}$~kpc. 

Note that this result relies on the assumption that we are tracing the motion of the same maser spot for all the 
epochs. It is known that the water maser emission of astronomical sources could be very variable, even on time scales 
of days. Particularly, the distribution of the masers in \k3 seems to have changed significantly since their discovery by 
Miranda~et~al.~(2001). However, we are confident that we are tracing the motion of the same maser spot in our 
VERA observations due to the following reasons: i) We are using the maser emission from the same velocity channel 
map for all the epochs (v$_{\rm LSR}\sim$22.57\kms). ii) The brightness of the maser spot used for the measurement 
did not change significantly during the observations (see Table~3). iii)~If we were not observing the same maser spot, 
the position would show sudden jumps as a function of time, instead of the smooth linear-sinusoidal motion seen in 
Figure~1; that would be reflected as data points that show significant departures from the fitted model. Therefore, even if 
the maser spot disappeared from one epoch to another, it would be tracing the same physical maser 
clump after reappearing. On the other hand, it is likely that the maser spot experienced non-linear motions during the 
two years of observations, slightly departing from the linear-sinusoidal fitted model. However, these departures are 
considered within the error bars shown in Figure~1. Thus, our estimation of the distance is not affected.

\input{table_5.tex}

\subsection{Internal proper motions of the water masers}

From our VLBA observations we mapped the maser emission towards the central region of \k3. 
The maser spots located at the tip of the lobes of \k3, as reported by Miranda et al. (2001),
 were not detected in our images above a 3--$\sigma$ noise level. The detected 
emission appears as individual maser features spread over an area 
of $\sim$20~$\times$~20 mas within a velocity range of v$_{\rm LSR}\sim$20~--~25~km~s$^{-1}$ 
and they are labeled as $A$, $B$, $C$, $D$, $E$, $F$ and $G$ in Figure 2. The integrated 
flux density of the strongest feature was approximately 20~Jy (see inset in left panel of Figure 2). 
This flux is higher than those observed by Miranda et al. (2001) and 
Tafoya et al. (2007) by a factor of $\sim$10. The features $A$, $B$, $E$, $F$ and $G$ were detected 
in the three epochs of observation, while the features $C$ and $D$ were only detected in 
two epochs (Table 4).  In general the emission from each feature appears in more than 3 channels (i.e. each 
maser feature consists of more than 3 maser spots). The relative position of the spots within a maser 
feature changes slightly from each other ($\sim$50 $\mu$as) at the different epochs, due to the uncertainty of 
the position and to the internal turbulent motion of the gas in the masing region. This variation is significant 
when compared to the internal proper motions  (see Table 5). Thus, following the same reasoning as in the 
case of the VERA observations (see \S 3.1), we have used the position of the strongest spot of 
each feature to determined their relative proper motions.  
 
\begin{figure*}[t]
\begin{center}
\includegraphics[angle=-0,scale=0.4]{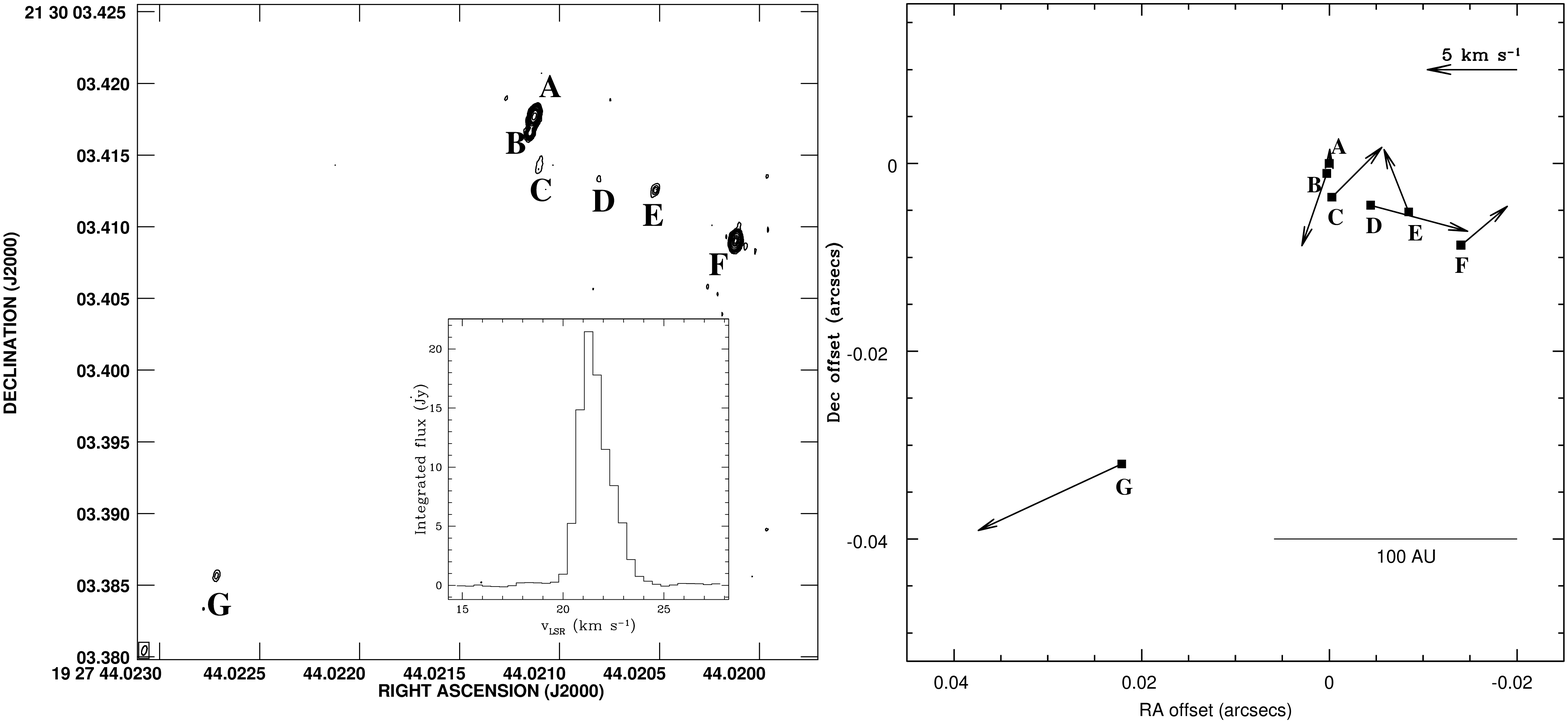}
\end{center}
\caption{Left panel: VLBA image of the water maser emission toward the central region of \k3. The image 
includes emission integrated in the velocity range from v$_{\rm LSR}=$ 15-28 \kms over an area of 
$\sim$20$\times$20 mas. The contours are ($-$3, $-$2, $-$1,1, 2, 3, 4, 5, 6, 7, 8, 9, 10, 20, 40, 60, 80, 
100, 200)$\times$0.085~Jy~beam$^{-1}$~km~s$^{-1}$. The peak of the image is 26.3 Jy~beam$^{-1}$~km~s$^{-1}$. 
The synthesized beam is  0.6$\times$0.3 mas (P.A.$=-$17$^{\circ}$) 
and is shown in the lower left corner of the image. Right panel: The proper motions of the masers in a 
frame of reference in which the average proper motion is zero. The upper arrow indicates the velocity of the masers 
using the distance to \k3 measured from the annual parallax obtained from the VERA observations.}
\label{fig:model-motions}
\end{figure*}

The spot used for the self-calibration process, which is contained within the feature labeled as 
$A$ in Figure 2, was initially chosen as the reference point to derive the relative proper motions of the other spots.
We measured the offsets and then performed a linear least squares fitting as a function of time. The 
result gives us the motions of the masers as seen from an observer attached to the feature $A$. However, since the absolute 
proper motion of the feature $F$ was measured from the VERA observations 
(see \S3.1), we decided to change the frame of reference of the relative proper motions from the feature $A$ to the feature 
$F$ by subtracting the corresponding proper motion vector (see Table 5). Subsequently, by adding the the absolute 
proper motion of the feature $F$ to the the relative proper motions of the other masers, we determined the absolute 
proper motions of all the features.

We noticed  that the absolute linear proper motion of the feature $F$ measured with VERA coincides with the axis of
 the bipolar outflow of \k3, suggesting that this maser, and the others as well, could be tracing 
the motion of the gas in the outflow. However, this proper motion is very similar to that of the source IRAS~19312+1950:  
 $\mu_{\alpha}=-$2.61~mas~yr$^{-1}$, $\mu_{\delta}=-$6.73~mas~yr$^{-1}$ (Imai et al. 2010).  This source 
 is located in the sky very close to \k3 ($\sim$2$^{\circ}$) at a very similar distance from the Sun $\sim$3.8 kpc).  
 This coincidence indicates that the proper motions of the masers in these two sources are due to their motion 
 within the Galaxy rather than to internal motions. Furthermore, in the case of \k3, it is reasonable to assume 
 that the magnitude of the internal motions of the masers projected on the plane of the sky is of the same order as 
 the radial velocity of the masers, v$_{\rm rad}$$\sim$5\kms~($\sim$0.3 mas~yr$^{-1}$), assuming a systemic velocity of 
 the source of $\sim$26\kms (Tafoya et al. 2007).  This value is one order of magnitude smaller than the 
 observed proper motions of the masers. Thus, we conclude that these absolute proper motions
 are mainly showing the motion of the whole nebula, which can be approximated as the average of all of them: 
  $\langle\mu_{\alpha}\rangle=-$3.34$\pm$0.10~mas~yr$^{-1}$, $\langle\mu_{\delta}\rangle=-$5.93$\pm$0.07~mas~yr$^{-1}$. 
More specifically, this proper motion corresponds to that of a frame of reference in which the sum of the motions 
of the masers equals zero.  The velocity vectors of the masers in that frame of reference are shown in the 
right panel of Figure 2 and the values of the proper motions are given in Table 5. 

\section{Discussion}

\subsection {Distance to \k3 and its location in our Galaxy}

The distance to \k3 has been commonly assumed to be $D\sim$5~kpc, as estimated from the distance 
scale proposed by Zhang (1995). This distance scale is based on the correlation between the ionized mass, 
the surface brightness temperature and the radii of the PNe. However, several other values have been 
estimated by other authors:  Cahn et al. (1992) calibrated the parameters of several PNe using the values from other PNe, 
for which the distances are well known, and estimated a distance to \k3 to be $D\sim$4~kpc. Aaquist  (1993) 
assumed that \k3 was associated to the L755 molecular cloud for which the kinematical distance based on the velocity of 
the CO is $D \sim$9 kpc. van de Steene \& Zijlstra (1995) obtained a distance of 6.38 kpc based on a 
correlation of the brightness temperature and size of the planetary nebulae. Phillips (2002) using the same correlation, 
but based upon nearby sources, obtained a distance for \k3 of 2.08 kpc. Later, Phillips (2004) re-examined his 
method and found a distance of  6.56 kpc. Recently, Giammanco et al. (2010) estimated the distance to PNe by means 
of extinction measurements; they found a distance to \k3, $D<$1~kpc. This value is much smaller than the previous 
ones and the one we present in this work.  This could be due to the fact that Giammanco et al. (2010) based their 
analysis on the extinction of the H$\alpha$ emission, which can be very uncertain for some PNe. This wide variety 
of values for the distance to \k3 shows how poorly it has been constrained by these methods. It also shows that while the 
observations that use some indirect technique to measure the distance to several PNe, in general, provide good 
statistical results (uncertainty of 50\% or less), the distance 
to individual objects, specially very obscured young PNe, might suffer from large uncertainties, in some cases 
by a factor of 2 or more.  This results in a considerable error when determining other parameters of the nebula. 
From our observations we have obtained the distance to \k3 to be $D\sim$3.9~kpc with an associated error of 
18\%. This result improves the accuracy of the value of  the distance to this source and has the advantage of 
being based on a direct measurement, without assumptions on the intrinsic parameters, extinction, etc. 
We emphasize that this is the first time that the distance to a PN is measure through the parallax of the masers.

In previous works related to \k3, such as those of Miranda et al. (2001); Tafoya et al. (2007); Uscanga et al. (2008); 
Gomez et al. (2009) in which several parameters of the nebula were obtained, the adopted distance was 5 kpc. 
Our result improves the estimation of the distance to a value that is  a factor of  $\sim$0.75 smaller than the value used 
by those authors. In consequence, it is necessary to introduce a correction factor of 0.75 for the values of those 
parameters that scale linearly with the distance, and a factor of $\sim$0.56, for those that scale as $\propto D^{2}$. 
Then, the masers associated with the bipolar ionized lobes, detected by Miranda et al. (2001), are located at  
$\sim$3,800 AU from the center of the nebula, which is also the extent of the ionized lobes. The projected distance 
to the masers located near the central region of the nebula is only $\sim$65 AU.  In general the time-scales depend 
linearly with the distance, which means that they should be shorter by a factor of 0.75 and that \k3 did not enter 
its PN phase until 1988. This result can explain why there was no He$^{+}$ emission from \k3 in 1986 (see Miranda 
et al. 2001 and references therein). On the other hand, the molecular mass of \k3 estimated by Tafoya et al. (2007) 
would change from 0.017$M_{\odot}$ to $\sim$0.01$M_{\odot}$, and the Zeeman pair found by G\'omez et al. (2009) 
would be located at around 110 AU, instead of 150 AU, from the central star.

Given our result on the parallax, it follows that \k3 is located at a distance of 3.9$^{+0.7}_{-0.5}$~kpc 
in the direction with Galactic longitude $l\sim$56$^{\circ}$, located 140$^{+25}_{-18}$ pc over the Galactic 
plane, as observed from our Solar System. On the other hand, if we assume that the distance from the Sun 
to the center of our Galaxy is $R_{\odot}=$ 8.5 kpc (Dehen \& Binney 1998), the galactocentric cylindrical coordinates of \k3 are: 
($R,\theta,z$)~=~(7.11$^{+0.08}_{-0.06}$ kpc, 27$\pm$5$^{\circ}$, 140$^{+25}_{-18}$ pc). Furthermore, we 
can estimate the motion of \k3 within our Galaxy by using our approximation of the proper motion discussed 
in \S 3.2. First, we use again the assumption that the distance from the Sun to the Galactic center is 
$R_{\odot}=$ 8.5 kpc; we also assume a Galactic rotation speed in the Solar circle, 
$\Theta_{\odot}=$ 220\kms, and a solar motion with respect to the Local Standard of Rest 
($U_{\odot},V_{\odot},W_{\odot}$)~=~(11.1, 12.24, 7.25) \kms~ (Sch\"onrich et al. 2010). Then, by following the 
formulae presented by Johnson \& Soderblom (1987), we obtain a velocity vector for \k3, 
($V_{R}, V_{\theta}, V_{z}$)~=~(33$\pm$16, 233$\pm$11, 11$\pm$2) \kms, with $V_{R}$ positive in 
the direction of the Galactic center. If we consider a flat rotation curve of the Galaxy, then the velocity of the Local 
Standard of Rest of \k3 is $\Theta_{\rm K 3-35}$$\sim$220\kms and the modulus of the deviation from circular 
motion is $\sim$37$\pm$14\kms. 

The derived values for the distance, height over the Galactic plane and peculiar motion of \k3 are similar 
to those found by Imai et al. (2010) for the water maser emitting post-AGB star IRAS 19312$+$1950, which 
lies only 2 degrees away from \k3. This suggests that they might be tracing a similar population of stars, 
i.e. relatively high-mass ($M_{\star}>$1.5$M_{\odot}$) evolved stars, located in the Galactic Thin Disk, and whose 
dynamical age, estimated from their velocity dispersion, is $\lesssim$1 Gyr (e. g. Soubrian et al. 2003). This would be in agreement 
with the idea that bipolar PN evolved from progenitors with relatively higher mass  
than the average PN (Corradi et al. 1995). However, to further understand the distribution and kinematics of 
evolved stars within our Galaxy, more measurements toward this type of stars, similar to those presented in this 
work, are required.

\subsection{The masers in the equatorial region}

As previously mentioned, Miranda et al. (2001) found maser emission toward three regions in \k3.
The maser emission at the tips of the lobes was suggested to be associated with a bipolar wind while 
the central masers were thought to be tracing an equatorial toroid. Uscanga et al.(2008) modeled the 
kinematical distribution of the water masers in the central region of this source using their radial velocity component. 
These authors found that the field of the radial velocity of the maser emission could be fitted by a model of 
a rotating and expanding ring-like structure tilted 55$^{\circ}$ with respect to the plane of the sky 
and with a position angle of 158$^{\circ}$. The expansion and rotation velocity of this ring is 1.4\kms 
and 3.1\kms, respectively. To try to test this model, we calculated and plotted the expected proper motions of the 
masers on the plane of the sky and compare them with those from our observations. Since we do not know 
exactly the velocity of the masers with respect to the central star, we plotted the proper motions of the masers 
with respect the the feature $F$. Then, we changed the frame of reference of the model to the maser with 
the same radial velocity as that of the feature $F$.  According to the model, the masers would move $\lesssim$1 
mas in 4 years, thus, the distribution of masers is not expected to change significantly. Assuming this, we 
can directly superimpose the masers and compare the velocity fields. We found that, while the magnitudes 
of the observed proper motions are similar to those of the model, the directions do not coincide. This disagreement 
could be due the fact that the masers presented by Miranda et al. (2001) could be arising in different maser clouds from 
those observed with the VLBA four years later. This idea is supported by the observations toward \k3 
presented by de~Gregorio-Monsalvo et al. (2004). These authors compared the positions of the water 
masers from observations carried out in May 2002 to the masers presented by Miranda et al. (2001) using 
the peak of the radio continuum as the point of reference. The separation of the masers between the two 
epochs implies expansion velocities of the masers larger than 100\kms~ in the equatorial region, which results 
unlikely. This separation can be better explained if we consider that \k3 is a very young planetary nebula whose 
ionized component is changing rapidly. Since the water maser emission is arising in the regions where the 
gas has not been ionized, the separation would be indicating the expansion of the ionization front,  which would 
have a speed $\sim$100\kms. 

\section{Summary}

We presented the results from observations of the water maser emission toward \k3 using two 
VLBI arrays: VERA and the VLBA. From the VERA observations we have measured the annual parallax 
of this source, yielding for the first time a direct measurement of its distance: $D\sim$3.9 kpc 
with an error of $\sim$18\%. The value of the distance is a factor of 0.75 smaller than the 
previously assumed value. This provides a correction factor of 0.75 for those parameters that scale linearly 
with the distance and a correction factor of 0.56  for those parameters that scale as $D^{2}$. 
This implies that this PN is younger than previously thought. From our measurements of the proper 
motion and distance, we also determine the location and kinematics of this source in our Galaxy. 
From the spatial distribution and relative internal motions of the masers, we suggest that the central region 
of \k3 is evolving rapidly as expected for a young PN and could be due to the expansion of the ionizing 
front and winds.

{\bf Acknowledgments}\\
D.T. acknowledges support from the Japan Society for Promotion of Science (project ID: 22-00022). H.I. has 
been financially supported by the Grant-in-Aid for Young Scientists from the Ministry of Education, 
Culture, Sports, Science, and Technology (18740109), and by the Grant-in-Aid for Scientific Research 
from Japan Society for Promotion Science (20540234). Y.G. acknowledges support from CONACyT grant 80769. 
G.A. and J.M.T. acknowledge support from MICINN (Spain) AYA2008-06189-C03 grant (co-funded with FEDER funds). 
and from Junta de Andaluc\'{\i}a (Spain). L.F.M. is supported partially by grants AYA2008-01934 of the Spanish 
MICINN (co-funded by FEDER funds) and FQM1747 of the Junta de Andaluc\'{\i}a. 

\end{document}

%% file: table_1.tex
\begin{table*}[t]

\begin{center}

\begin{minipage}{0.65\textwidth}

\caption{Parameters of the VERA observations.}\label{tab:status}
\footnotesize

\begin{tabular}{c@{\hspace{1cm}}l@{\hspace{0.4cm} }lcc@{ }c} 
\hline

\multicolumn{6}{c}{}\\[-2ex]
\multicolumn{2}{c}{Observation\phantom{aaa}}  & 
\multicolumn{2}{c}{VERA}         & 
\multicolumn{1}{c}{Beam\footnotemark[c]}     & 

\multicolumn{1}{c}{Astrometry}  \\

\multicolumn{1}{l}{\phantom{a}Code}         & 
\multicolumn{1}{l}{Epoch}        & 
\multicolumn{1}{l}{\hspace{-0.3cm}Telescopes\footnotemark[a]}& 
\multicolumn{1}{c}{Noise\footnotemark[b]}    &  
\multicolumn{1}{c}{[mas,\hspace{2mm}$^{\circ}$]}  & 
\multicolumn{1}{c}{valid?}       \\ 

\hline 
r07298a & 2007 October 25 & MROS & 56& 1.34$\times$0.78$\;$, $-$45.8 & Yes \\
r07333a & 2007 November 29 & MROS & 78   & 1.45$\times$0.70$\;$, $-$50.7 & Yes\\
r07358b & 2007 December 25 & MROS & 82  & 1.49$\times$0.53$\;$, $-$50.8 & Yes \\ 
r08042b & 2008 February 11 & MROS & 40  & 1.29$\times$0.88$\;$, $-$57.3 & Yes \\ 
r08080b & 2008 March 20 & MROS & 38  & 1.34$\times$0.83$\;$, $-$47.5 & Yes \\ 
r08106a & 2008 April 15 & MROS & 66& 1.17$\times$0.63$\;$, $-$55.9 & Yes \\ 
r08142a & 2008 May 21 & MROS & 61  & 1.77$\times$0.80$\;$, $-$45.3 & No\footnotemark[d] \\ 
r08182a & 2008 June 30 & MROS & 87  & 1.44$\times$0.74$\;$, $-$50.4 & No\footnotemark[d] \\ 
r08211a & 2008 July 29 & MROS & 127 & 1.31$\times$0.77$\;$, $-$52.3 & No\footnotemark[d] \\ 
r08271a & 2008 September 27 & MO & --- & \multicolumn{1}{c}{---} & No\footnotemark[e]\\ 
r08344a & 2008 December 11 & MROS & 65  & 1.48$\times$0.78$\;$, $-$55.4 & Yes \\ 
r09013b & 2009 January 13 & MROS & 41  & 1.23$\times$0.73$\;$, $-$47.8 & Yes \\ 
r09045a & 2009 February 14 & MRS & 94  & 1.95$\times$0.73$\;$, $-$44.5 & No\footnotemark[f] \\ 
r09126a & 2009 May 8 & MROS & 54  & 1.75$\times$0.77$\;$, $-$47.0 & No\footnotemark[d]\\ 
r09160a & 2009 June 11 & MROS & 93  & 1.41$\times$0.75$\;$, $-$49.6 & Yes \\ 
r09241a & 2009 August 29 & MROS & 79  & 1.62$\times$0.73$\;$, $-$45.2 & No\footnotemark[g]\\
\hline 
\end{tabular}

\footnotetext[a]{Telescope whose data were valid for phase-referencing maser imaging. M: Mizusawa, R: Iriki, O: Ogasawara, S: Ishigakijima.}
\footnotetext[b]{rms noise in units of mJy beam$^{-1}$ in the emission-free spectral channel image.}
\footnotetext[c]{Synthesized beam size resulting from natural weighted visibilities; major and minor axis lengths and position angle.}
\footnotetext[d]{Possible large position offsets due to a systematic error.}
\footnotetext[e]{Visibility data valid for the image synthesis were obtained only from the Mizusawa and Ogasawara stations.}
\footnotetext[f]{Visibility data valid for the image synthesis were obtained only from the Mizusawa,  Iriki, and Ishigakijima stations.}
\footnotetext[g]{{\bf No maser emission was detected.}}
\end{minipage}
\end{center}
\end{table*}

%% file: table_2.tex
\begin{center}
\footnotesize{
\begin{longtable}[t!]{l@{\hspace{0cm}}rl@{\hspace{0.3cm}}rl@{\hspace{0.2cm}}cc}

\caption{Parameters of the H$_{2}$O maser spots in \k3 detected with VERA} \label{grid_mlmmh} \\
\hline 
\multicolumn{7}{c}{\phantom{.}\hspace{6cm}\phantom{.}}\\[-2ex]
\multicolumn{1}{c}{v$_{\rm LSR}$\footnotemark[h]} & 
\multicolumn{2}{c}{R.A. offset~~($\sigma$)\footnotemark[i]} & 
\multicolumn{2}{c}{Dec offset~~($\sigma$)\footnotemark[i]} & 
\multicolumn{1}{c}{$I$\footnotemark[j]}&
\multicolumn{1}{c}{Feauture\footnotemark[k]} \\
\multicolumn{1}{c}{[km s$^{-1}$]} & 
\multicolumn{2}{c}{[mas]} & 
\multicolumn{2}{c}{[mas]} & 
\multicolumn{1}{c}{[Jy beam$^{-1}$]} &
\multicolumn{1}{c}{label} \\ 
\hline 
\endfirsthead

\multicolumn{7}{c}%
{{\tablename\ \thetable{} -- continued from previous page}} \\
\hline 
\multicolumn{7}{c}{}\\[-2ex]
\multicolumn{1}{c}{v$_{\rm LSR}$\footnotemark[h]} & 
\multicolumn{2}{c}{R.A. offset~~($\sigma$)\footnotemark[i]} & 
\multicolumn{2}{c}{Dec offset~~($\sigma$)\footnotemark[i]} & 
\multicolumn{1}{c}{$I$\footnotemark[j]}&
\multicolumn{1}{c}{Feauture\footnotemark[k]} \\
\multicolumn{1}{c}{[km s$^{-1}$]} & 
\multicolumn{2}{c}{[mas]} & 
\multicolumn{2}{c}{[mas]} & 
\multicolumn{1}{c}{[Jy beam$^{-1}$]} &
\multicolumn{1}{c}{label} \\ 

\hline 

\endhead

\multicolumn{7}{c}{}\\[-2ex]
\multicolumn{7}{c}{{Continued on next page}} \\ \hline
\endfoot

\hline 

\multicolumn{7}{c}{}\\[-2ex]
\multicolumn{7}{p{7cm}}{\parbox[l]{10.5cm}{\footnotemark[h]\hspace{3mm}Local-Standar-of-Rest velocity {\bf of the channel map}.}} \\
\multicolumn{7}{p{7cm}}{\parbox[l]{10.5cm}{\footnotemark[i]\hspace{3mm}Position offset with respect to the delay-traacking center of the data co\-rrelation (see main text) in units of mas. The value in parenthesis is the relative position error given by the AIPS task JMFIT.}} \\
\multicolumn{7}{p{7cm}}{\parbox[l]{10.5cm}{\footnotemark[j]\hspace{3mm}Peak intensity of the maser spot.}}\\
\multicolumn{7}{p{7cm}}{\parbox[l]{10.5cm}{\footnotemark[k]\hspace{3mm}{\bf Label of the maser feature to which the maser spot belongs (see main text and Fig.~1). No label was assigned for those maser spots that did not show association with the other 
maser feature and that only appeared sporadically.}}}
\endlastfoot

%

\multicolumn{7}{c}{2007 October 25} \\ \hline
    22.57&     $-$49.86&   (0.02) &     $-$36.16&   (0.02) &         1.70 & F\\
    22.15&     $-$49.84&   (0.02) &     $-$36.11&   (0.02) &         1.41 & F\\
    21.30&     $-$35.37&   (0.01) &     $-$27.48&   (0.02) &         2.32 & A\\
    20.88&     $-$35.32&   (0.01) &     $-$27.43&   (0.01) &         2.34 & A\\
\hline
\multicolumn{7}{c}{2007 November 29} \\ \hline
    22.57&     $-$50.07&   (0.01)  &     $-$36.85&   (0.01) &         2.87 & F\\
    22.15&     $-$50.03&   (0.01)  &     $-$36.86&   (0.01) &         2.80 & F\\
    21.73&     $-$35.62&   (0.01)  &     $-$28.26&   (0.02) &         2.89 & A\\
    21.30&     $-$35.61&   (0.01)  &     $-$28.23&   (0.01) &        12.20 & A\\
    20.88&     $-$35.61&   (0.01)  &     $-$28.21&   (0.01) &         8.02 & A\\
    20.46&     $-$35.63&   (0.03)  &     $-$28.06&   (0.04) &         0.67 & A\\
\hline
\multicolumn{7}{c}{2007 December 25} \\ \hline
    22.57&     $-$51.20&   (0.07)  &     $-$39.98&   (0.04)  &         0.89  & --\\
    22.57&     $-$50.26&   (0.04)  &     $-$37.24&   (0.03)  &         1.59  & F\\
    22.15&     $-$50.27&   (0.05)  &     $-$37.24&   (0.04)  &         0.92  & F\\
    22.15&     $-$48.71&   (0.07)  &     $-$37.06&   (0.06)  &         0.68  & --\\
    21.73&     $-$35.79&   (0.05)  &     $-$28.66&   (0.04)  &         0.77  & A\\
    21.30&     $-$35.75&   (0.02)  &     $-$28.67&   (0.02)  &         4.04  & A\\
    20.88&     $-$35.78&   (0.03)  &     $-$28.65&   (0.03) &          3.23  & A\\
    21.30&     $-$34.32&   (0.06)  &     $-$28.44&   (0.04) &          1.47  & --\\
\hline
\multicolumn{7}{c}{2008 February 11} \\ \hline
    22.99&     $-$50.36&   (0.03)  &     $-$37.80&   (0.03)  &         0.57  & F\\
    22.57&     $-$50.38&   (0.01)  &     $-$37.81&   (0.01)  &         2.98  & F\\
    22.15&     $-$50.39&   (0.01)  &     $-$37.79&   (0.01)  &         1.95  & F\\
    21.73&     $-$35.82&   (0.03)  &     $-$30.14&   (0.03)  &         0.75  & A\\
    21.30&     $-$35.81&   (0.03)  &     $-$29.62&   (0.05)  &         0.66  & A\\
    20.88&     $-$35.88&   (0.03)  &     $-$29.12&   (0.02)  &         0.70  & A\\
\hline									     
\multicolumn{7}{c}{2008 March 20} \\ \hline				     
    22.57&     $-$50.68&   (0.02) &     $-$38.55&   (0.02)  &         2.62 & F\\
    22.15&     $-$50.68&   (0.02) &     $-$38.53&   (0.02)  &         1.32 & F\\
    21.30&     $-$36.16&   (0.03) &     $-$30.03&   (0.03)  &         0.79 & A\\
    20.88&     $-$36.18&   (0.02) &     $-$29.85&   (0.02)  &         0.74 & A\\
\hline						    			   
\multicolumn{7}{c}{2008 April 15} \\ \hline	    			   
    22.57&     $-$51.15&   (0.01) &     $-$38.83&   (0.01) &         2.16 & F\\
    22.15&     $-$51.11&   (0.02) &     $-$38.88&   (0.02) &         1.19 & F\\
    21.30&     $-$36.65&   (0.01) &     $-$30.22&   (0.01) &         2.03 & A\\
    20.88&     $-$36.69&   (0.02) &     $-$30.15&   (0.02) &         1.51 & A\\
\hline			        	       	     			  
\multicolumn{7}{c}{2008 May 21} \\ \hline      	     			  
    22.57&     $-$51.59&   (0.09) &     $-$39.23&   (0.11) &         2.52 & F\\
    22.15&     $-$51.58&   (0.09) &     $-$39.24&   (0.11) &         1.52 & F\\
    20.88&     $-$36.98&   (0.08) &     $-$30.66&   (0.10) &         1.17 & A\\
\hline			        	       	     
\multicolumn{7}{c}{2008 June 30} \\ \hline     	     
    22.57&     $-$52.34&   (0.04) &     $-$39.24&   (0.04) &         1.75 & F\\
    22.15&     $-$52.70&   (0.10) &     $-$42.78&   (0.09) &         0.62 & --\\
    22.57&     $-$50.89&   (0.05) &     $-$41.17&   (0.05) &         1.07 & --\\
    22.15&     $-$52.33&   (0.05) &     $-$39.29&   (0.06) &         0.73 & F\\
    21.30&     $-$37.88&   (0.04) &     $-$30.69&   (0.05) &         0.80 & A\\
\hline			 		       				  
\multicolumn{7}{c}{2008 July 29} \\ \hline				  
    22.57&     $-$50.55&   (0.08) &     $-$38.30&   (0.08) &         0.87 & --\\
    22.57&     $-$51.42&   (0.13) &     $-$34.88&   (0.07) &         0.72 & --\\
\hline						    
\multicolumn{7}{c}{2008 December 11} \\ \hline	    
    22.99&     $-$53.59&   (0.03) &     $-$42.84&   (0.02) &         0.72 & F\\
    22.57&     $-$44.91&   (0.05) &     $-$39.31&   (0.04) &         0.45 & H\\
    22.57&     $-$53.63&   (0.03) &     $-$42.79&   (0.02) &         0.81 & F\\
    22.15&     $-$44.95&   (0.05) &     $-$39.30&   (0.05) &         0.35 & H\\
    22.15&     $-$53.62&   (0.09) &     $-$42.85&   (0.06) &         0.47 & F\\
\hline			        	      	     
\multicolumn{7}{c}{2009 January 13} \\ \hline	     
    22.99&     $-$45.00&   (0.04) &     $-$40.17&   (0.04) &         0.39 & H\\
    22.99&     $-$53.78&   (0.02) &     $-$43.58&   (0.02) &         0.62 & F\\
    22.57&     $-$45.04&   (0.02) &     $-$40.12&   (0.02) &         0.79 & H\\
    22.57&     $-$53.80&   (0.02) &     $-$43.54&   (0.02) &         0.73 & F\\
    22.15&     $-$45.07&   (0.03) &     $-$40.09&   (0.03) &         0.41 & H\\
    22.15&     $-$53.78&   (0.04) &     $-$43.56&   (0.04) &         0.26 & F\\
\hline			      			     			  
\multicolumn{7}{c}{2009 February 14} \\ \hline	     			  
    22.57&     $-$54.37&   (0.03) &     $-$43.51&   (0.03) &         1.13 & F\\
    22.57&     $-$52.63&   (0.09) &     $-$41.70&   (0.09) &         0.46 & --\\
    22.15&     $-$45.62&   (0.05) &     $-$40.14&   (0.05) &         0.62 & H\\
    22.15&     $-$54.21&   (0.06) &     $-$43.64&   (0.06) &         0.66 & F\\
    21.30&     $-$39.77&   (0.07) &     $-$35.59&   (0.08) &         0.42 & A\\
\hline			        	      	     			  
\multicolumn{7}{c}{2009 May 8} \\ \hline	     			  
    22.57&     $-$55.37&   (0.07) &     $-$44.54&   (0.09) &         0.75 & F\\
    22.15&     $-$55.31&   (0.09) &     $-$44.60&   (0.09) &         0.44 & F\\
\hline
\multicolumn{7}{c}{2009 June 11} \\ \hline
    22.57&     $-$55.41&   (0.09) &     $-$45.49&   (0.09) &         0.63 & F\\
    22.57&     $-$55.14&   (0.06) &     $-$45.30&   (0.07) &         0.70 & F\\
\end{longtable}

}

\end{center}

%% file: table_3.tex
\begin{table*}[t]

\footnotesize
\begin{center}
\begin{minipage}{.65\hsize}

\caption{Data obtained with VERA used to perform the fitting of the motion of the maser 
emission in the channel at v$_{\rm LSR}$=22.57 {\bf km~s$^{-1}$ (i.e. $m_{\rm 22.57}$, see the main text).}} \label{tab:astrometry}

\end{minipage}
\end{center}

\begin{center}
\begin{tabular}{@{\hspace{0.5cm}}l@{\hspace{0.5cm}}cc@{\hspace{0.3cm}}cc@{\hspace{0.3cm}}c@{\hspace{0.2cm}}} \hline 

\multicolumn{1}{c}{Epoch}         & 
\multicolumn{1}{c}{R.A. offset}        & 
\multicolumn{1}{c}{($\sigma$)}& 
\multicolumn{1}{c}{Dec. offset}    &  
\multicolumn{1}{c}{($\sigma$)} &
\multicolumn{1}{c}{$I$}\\
\multicolumn{1}{c}{}         & 
\multicolumn{1}{c}{[mas]}        & 
\multicolumn{1}{c}{[mas]}& 
\multicolumn{1}{c}{[mas]}    &  
\multicolumn{1}{c}{[mas]} &
\multicolumn{1}{c}{[Jy beam$^{-1}$]}    \\
\hline
                    
2007 October 25   & $-$49.86 & (0.02) & $-$36.16 & (0.02)&1.70\\  
2007 November 29  & $-$50.07 & (0.01) & $-$36.85 & (0.01)& 2.87\\  
2007 December 25  & $-$50.26 & (0.04) & $-$37.24 & (0.03)& 1.59\\  
2008 February 11  & $-$50.38 & (0.01) & $-$37.81 & (0.01)& 2.98\\  
2008 March 20     & $-$50.68 & (0.02) & $-$38.55 & (0.02)& 2.62\\  
2008 April 15     & $-$51.15 & (0.01) & $-$38.83 & (0.01)& 2.16\\  
2008 December 11  & $-$53.63 & (0.03) & $-$42.79 & (0.02)& 0.81\\  
2009 January 13   & $-$53.80 & (0.02) & $-$43.54 & (0.02)& 0.73\\  
2009 June 11 	  & $-$55.14 & (0.06) & $-$45.30 & (0.07)& 0.70\\
\hline

\end{tabular}

\end{center}

\end{table*}

%% file: table_4.tex
\begin{table*}[t]
\footnotesize

\begin{center}
\begin{minipage}{0.7\hsize}

\caption{Positions of water maser spots in K 3-35 from the VLBA observations\footnotemark[l]}\label{table_4}

\begin{tabular}{c@{\hspace{1cm}}crr@{\hspace{1cm}}rrrr}\hline

\multicolumn{1}{c}{maser}&
\multicolumn{1}{c}{v$_{\rm LSR}$}&
\multicolumn{1}{c}{RA offset\footnotemark[m]}&
\multicolumn{1}{l}{($\sigma$)\footnotemark[n]}&
\multicolumn{1}{c}{Dec offset\footnotemark[m]}&
\multicolumn{1}{l}{($\sigma$)\footnotemark[n]}&
\multicolumn{1}{l}{I}	&
\multicolumn{1}{l}{($\sigma$)}\\

\multicolumn{1}{c}{}&
\multicolumn{1}{c}{km~s$^{-1}$}&
\multicolumn{1}{c}{[mas]}&
\multicolumn{1}{l}{[mas]}&
\multicolumn{1}{c}{[mas]}&
\multicolumn{1}{l}{[mas]}&
\multicolumn{1}{l}{[Jy]}&	
\multicolumn{1}{l}{[Jy]}\\
\hline
     
\multicolumn{8}{c}{1st Epoch: } \\ \hline
B &  21.26 &0.261  &  (0.001) &  $-$1.062  & (0.002)&0.674&(0.006)\\
C & 20.84 &$-$0.279  &  (0.004) &  $-$3.598  & (0.008)&0.096&(0.005)\\
D & 22.10 &$-$4.406  &  (0.014) &  $-$4.446  & (0.023)&0.035&(0.005)\\
E & 22.10 &$-$8.464  &  (0.003) &  $-$5.155  & (0.004)&0.192&(0.005)\\
F & 22.52 &$-$14.030  &  (0.001) &  $-$8.705  & (0.001)&2.396&(0.006)\\
G &  23.36 & 22.110  &  (0.034) &  $-$32.000  & (0.040)&0.013&(0.004)\\
\hline
\multicolumn{8}{c}{2nd Epoch} \\ \hline
B & 21.26& 0.276  &  (0.001)  & $-$1.098  &(0.002)&1.004&(0.009)\\
C & 20.84&$-$0.303  &  (0.008)  & $-$3.581  &(0.015)&0.155&(0.011)\\
D & 22.10&$-$4.452  &  (0.015)  & $-$4.465  &(0.023)&0.074&(0.011)\\
E & 22.10&$-$8.451  &  (0.004)  & $-$5.131  &(0.006)&0.291&(0.011)\\
F & 22.52&$-$14.050  &  (0.001)  & $-$8.692  &(0.001)&6.960&(0.011)\\
G & 23.36& 22.170  &  (0.005)  & $-$32.040  &(0.009)&0.128&(0.008)\\
\hline
\multicolumn{8}{c}{3rd Epoch} \\ \hline
B &  21.26 &0.277   & (0.001)  & $-$1.126   & (0.001)&1.099&(0.010)\\
E & 22.10 &$-$8.449   & (0.009)  & $-$5.121   & (0.009)&0.143&(0.008)\\
F &  22.52&$-$14.070   & (0.001)  & $-$8.688   & (0.001)&3.143&(0.009)\\
G & 23.36 & 22.210   & (0.014)  & $-$32.060   & (0.020)&0.064&(0.007)\\
\hline
\end{tabular}

\footnotetext[l]{The synthesized beam is  0.6$\times$0.3 mas (P.A.=$-$17$^{\circ}$)}
\footnotetext[m]{The positions are referred to the brightest spot labeled as $A$ in Figure 2.}
\footnotetext[n]{Relative position errors.}

\end{minipage}
\end{center}

\end{table*}

%% file: table_5.tex
\begin{table*}[h]
\footnotesize

\begin{center}
\begin{minipage}{0.55\hsize}

\caption{Relative proper motions of the water maser in K 3$-$35}\label{table_5}

\begin{tabular}{crrrr}\hline

\multicolumn{1}{c}{maser}&
\multicolumn{1}{c}{$\mu_{\alpha}$~($\sigma$)\footnotemark[o]}&
\multicolumn{1}{c}{$\mu_{\delta}$~($\sigma$)\footnotemark[o]}&
\multicolumn{1}{c}{$\mu_{\alpha}$~($\sigma$)\footnotemark[p]}&
\multicolumn{1}{c}{$\mu_{\delta}$~($\sigma$)\footnotemark[p]}\\

\multicolumn{1}{c}{}&
\multicolumn{1}{c}{[mas yr$^{-1}$]}&
\multicolumn{1}{c}{[mas yr$^{-1}$]}&
\multicolumn{1}{c}{[mas yr$^{-1}$]}&
\multicolumn{1}{c}{[mas yr$^{-1}$]}\\

\hline
$A$  & 0.140 (0.025)   &  $-$0.075   (0.010)   &$-$0.001  (0.088)  &   0.043  (0.042)   \\ 
$B$  & 0.217 (0.031)          &  $-$0.337  (0.027) & 0.076     (0.090)    & $-$0.220     (0.049)\\
$C$        &  $-$0.011 (0.025)      &   0.034  (0.010)&   $-$0.152    (0.088)   &   0.151  (0.042)\\ 
$D$    & $-$0.154 (0.025)       &   $-$0.197  (0.010) &   $-$0.295   (0.088) &   $-$0.079   ( 0.042)\\
$E$      & 0.216 (0.030)          &   0.071  (0.013)   &0.075     (0.089) &      0.189    (0.042) \\
$F$       & $\dots$\phantom{00000} & $\dots$\phantom{00000}&  $-$0.141    ( 0.092) &     0.118     (0.043) \\
$G$      & 0.578 (0.047)          &  $-$0.321  (0.013)   & 0.437     (0.097)   &  $-$0.203     (0.043)\\
\hline
\end{tabular}

\footnotetext[o]{The proper motions are referred to the brightest spot labeled as $F$.}
\footnotetext[p]{The proper motions are referred to a reference frame in which the sum of the proper motions equals zero.}

\end{minipage} 
\end{center}

\end{table*}